\documentclass[letterpaper,english,reprint, aps]{revtex4-2}
\usepackage[T1]{fontenc}
\usepackage[latin9]{inputenc}
\usepackage{amsmath}
\usepackage{amssymb}
\usepackage{stackrel}
\usepackage{graphicx}
\usepackage{esint}



\usepackage{babel}
\begin{document}
\preprint{APS/123-QED}
\title{Transition from localized to delocalized trajectories in random walk subject to random drives}
\author{Zijun Li$^{*}$}
\author{Jiming Yang}
\thanks{Equal contribution.}
\affiliation{Zhixin High School, Guangzhou, Guangdong, China }
\author{Huiyu Li}
\affiliation{Atlas Science, Princeton, New Jersey, USA}

\date{\today}
\begin{abstract}
Random walk subject to random drive has been extensively employed as a model for physical and biological processes. While equilibrium statistical physics has yielded significant insights into the distributions of dynamical fixed points of such a system, its non-equilibrium properties remain largely unexplored. In contrast, most real-world applications concern the dynamical aspects of this model. In particular, dynamical quantities like heat dissipation and work absorption play a central role in predicting and controlling non-equilibrium phases of matter. Recent advances in non-equilibrium statistical physics enable a more refined study of the dynamical aspects of random walk under random drives. We perform a numerical study on this model and demonstrate that it exhibits two distinct phases: a localized phase where typical random walk trajectories are non-extensive and confined to the neighborhood of fixed points, and a delocalized phase where typical random walk trajectories are extensive and can transition between fixed points. We propose different summary statistics for the heat dissipation and show that these two phases are distinctly different. Our characterization of these distinctive phases deepens the understanding of and provides novel strategies for the non-equilibrium phase of this model.

\end{abstract}
\keywords{Dynamical systems, Statistical mechanics}
\maketitle

\section{Introduction}

Recent advances in non-equilibrium statistical physics \citep{jarzynski1997nonequilibrium, crooks1999entropy, seifert2012stochastic} have shifted the focus from properties of equilibrium states to the thermodynamic analysis of transient properties in a system's history. Heat dissipation, or entropy production, has provided significant insights into the understanding and control of non-equilibrium phases of matter, with notable applications in biological processes such as self-assembly and self-replication \citep{murugan2015multifarious, zhong2017associative, bisker2018nonequilibrium, england2013statistical, baiesi2018life, floyd2019quantifying, perunov2016statistical}. It has been hypothesized that, given the same physical configuration space, living systems are more likely to exhibit high dissipation trajectories compared to non-living systems. Such self-organization phenomena, named \textit{dissipative adaptation} \citep{england2015dissipative}, have been studied in the context of many-body systems driven out-of-equilibrium \citep{horowitz2017spontaneous, kedia2019drive, gold2019self, zhong2020quantifying, zhong2021machine}. Therefore, understanding and quantifying physical conditions that lead to high dissipation is important. In this paper, we study dissipation in the model of random walk subject to random drives, a model extensively used for diffusion processes \citep{bouchaud1990classical, bouchaud1990anomalous}, active matter \citep{romanczuk2012active, bechinger2016active, cates2015motility}, and directed movements in biology \citep{codling2008random, bialek2012biophysics}. While equilibrium statistical physics has provided significant insights into the distributions of dynamical fixed points of such a system \citep{bouchaud1990anomalous}, its non-equilibrium properties remain largely unexplored. 

Through a detailed numerical analysis, we uncover two contrasting phases within this model: a localized phase, characterized by non-extensive, confined random walk paths near fixed points, and a delocalized phase, marked by extensive random walk paths with the ability to move between fixed points. Similar localization-delocalization transitions have been discussed in the context of large deviation theory in various settings, from random media \citep{solomon1975random, greven1994large} to random graphs \citep{coghi2019large, carugno2023delocalization, de2016rare, stuhrmann2023understanding} and quantum systems \citep{marino2022dynamical, smale2019observation}. We introduce various summary statistics to analyze heat dissipation and establish the distinct nature of these two phases. This exploration provides a novel numerical strategy to characterize the localization-delocalization phase transition in this model, and offers innovative approaches for studying its non-equilibrium behavior.

Our paper is organized as follows: In Section \ref{sec:Langevin-system} we introduce the Langevin equation for random walk and motivate the model for its random drive. In Section \ref{sec:Massive-vector-field} we introduce the vector field model for the random drive and discuss the Metropolis algorithm for simulating it. In Section \ref{sec:Langevin+MVF} we combine the Langevin equation and vector field to study random walk under random drive and discuss different regimes of the model. Finally, in Section \ref{sec:dissipation} we propose different metrics for detecting these two phases and demonstrate that the behavior of these metrics is distinctly different in the two phases.

\section{Langevin system\label{sec:Langevin-system}}

The original Langevin equation \citep{uhlenbeck1930theory} describes
Brownian motion, the apparently random movement of a particle in a
fluid due to collisions with the molecules of the fluid,
\begin{equation}
m\frac{dv}{dt}=-\gamma v+\eta(t).
\end{equation}

Here, $v$ is the velocity of the particle, and $m$ is its mass.
The force acting on the particle is written as a sum of a viscous
force proportional to the particle's velocity (Stokes\textquoteright s
law), and a noise term $\eta(t)$ representing the effect of the collisions
with the molecules of the fluid. The force $\eta(t)$ has a Gaussian
probability distribution with the correlation function 
\begin{equation}
\left\langle \eta_{i}(t)\eta_{j}(t')\right\rangle =2\gamma k_{B}T\delta_{ij}\delta(t-t').
\end{equation}

However, the Langevin equation is used to describe the motion of a
\textquotedbl macroscopic\textquotedbl{} particle at a much longer
time scale. In our research, we are going to study the Langevin model
in the regime that particles are moving in an extremely sticky fluid
with low Reynold's number. In this limit, we can consider the system
as a non-inertial system. In such non-inertial system, acceleration
can be neglected. In this limit, the original Langevin equation is
transformed into
\begin{equation}
-\gamma\dot{\vec{x}}+F(\vec{x})+\eta(t)=0.\label{eq:langevin1}
\end{equation}

Here, Eq.\ref{eq:langevin1} describes a general overdamped Langevin
system with many degrees of freedom $\vec{x}$. $F(\vec{x})$ models
the complicated external drivings that depends on the coordinate $\vec{x}$.
We can further decompose $F$ into two components, $F(\vec{x})=F_{ext}(\vec{x})-\nabla U_{int}(\vec{x})$.
$F_{ext}(\vec{x})$ is the external drive force that the particle
experiences, and $U_{int}(\vec{x})$ is the internal potential that
the particles live in and models the internal dynamics of the system.
For simplicity, in the following we consider the case with $U_{int}(\dot{\vec{x}})=0$
and focus on the effect of the external drive. Therefore, the Langevin
system of interest is
\begin{equation}
\gamma\dot{\vec{x}}=F_{ext}(\vec{x})+\eta(t).\label{eq:langevin2}
\end{equation}

In many applications of the model $\ref{eq:langevin2},F_{ext}$ is
most relevant when it models a random external drive with strong correlation
within nearby coordinates $\vec{x}$. In general, we require that
\begin{equation}
\left\langle F_{ext}(\vec{x})F_{ext}(\vec{x}')\right\rangle \sim e^{-\frac{|\vec{x}-\vec{x}'|}{\xi}},\label{eq:correlation}
\end{equation}
 where $\xi$ is the correlation length within the model. 

\section{Massive vector field\label{sec:Massive-vector-field}}

Eq.\ref{eq:correlation} is observed in many statistical models away
from critical points. In particular, we propose to use the massive
vector field, which is a much studied model in both high energy physics
\citep{srednicki2007quantum} and statistical physics \citep{kardar2007statistical}.
It is also widely employed as model of phase transition in the Ginzburg-Landau
theory.

\subsection{The model}

We consider a system of vectors $\vec{A}$ in a flat background with
the Hamiltonian:

\begin{equation}
H[\vec{A}]=\frac{1}{2}J|\nabla\vec{A}|^{2}+\frac{1}{2}m^{2}|\vec{A}|^{2}.\label{eq:mvf_hamiltonian}
\end{equation}
$H[\vec{A}]$, the Hamiltonian, describes the total energy of the
vector field. The first term stands for its kinetic energy, which
measures the cooperativity of the vectors, and $J$ describes the
interaction strength. The second term is called the mass term, in
which $m$ stands for the mass of the vectors, which measures the
randomness of the vectors. To see this, note that in the limit when
$J=0$, the equilibrium distribution of Eq.\ref{eq:mvf_hamiltonian}
is 

\begin{equation}
p[\vec{A}]|_{J=0}\sim\exp\left\{ -\beta H[\vec{A}]\right\} =\exp\left\{ -\frac{1}{2}m^{2}|\vec{A}|^{2}\right\} ,
\end{equation}

where $\beta$ is the inverse temperature. To gain more physical intuition,
we can compare our vector field model with the XY model\citep{stanley1968dependence,chaikin1995principles}.
XY model is a $n$-dimensional lattice. On each lattice site there
is an unit-length vector $\vec{S}_{i}=(\sin\theta_{i},\cos\theta_{i})$.
The Hamiltonian is given by 

\begin{equation}
H=-\frac{1}{2}J\stackrel[<i,j>]{\infty}{\sum}\vec{S_{i}}\cdot\vec{S_{j}}=-\frac{1}{2}J\stackrel[<i,j>]{\infty}{\sum}\cos(\theta_{i}-\theta_{j})\label{eq:XY}
\end{equation}
as $\mid\vec{S_{i}}\mid=\text{\ensuremath{\mid\vec{S_{j}}\mid}}=1$
and $J$ stands for the cooperativity between vectors. Our massive
vector field model Eq.\ref{eq:mvf_hamiltonian} can be thought of
as the continuum limit of Eq.\ref{eq:XY} with fixed vector length
$|\vec{A}|=1$ and variable angles. 

\subsection{Simulation Method}

In order to obtain the equilibrium configuration of Eqn.\ref{eq:mvf_hamiltonian},
we use Metropolis\textendash Hastings algorithm\citep{hastings1970monte}.
Metropolis\textendash Hastings algorithm use Monte-carlo methods to
obtain a sequence of random samples from a probability distribution
when direct sampling is difficult. This sequence can be used to approximate
the distribution. In our case, the distribution of interest is the
equilibrium distribution that we would like to sample from. We focus
on the case of 2-dimension because the simulation is relatively simple
and the configurations are already highly non-trivial. We decompose
the vector field into its horizontal and vertical components, $\vec{A}=(A_{h},A_{v})$,
such that $\mid\vec{A}\mid^{2}=\mid A_{h}\mid^{2}+\mid A_{v}\mid^{2}$.
Now Eq.\ref{eq:mvf_hamiltonian} can be written as

\begin{equation}
H[\vec{A}]=\frac{1}{2}J(\mid\nabla A_{h}\mid^{2}+\mid\nabla A_{v}\mid^{2})+\frac{1}{2}m^{2}(\mid A_{h}\mid^{2}+\mid A_{v}\mid^{2})
\end{equation}

We start our simulation with a normally-distributed initial conditions
$A_{v}$ and $A_{h}$ on every lattice site, and we assume periodic
boundary condition. 

In Metropolis-Hastings algorithm, the probability of observing configuration
$\vec{A}$ is assumed to be Boltzmann distributed.

\begin{equation}
P(\vec{A})=\frac{e^{-\beta H[\vec{A}]}}{Z}
\end{equation}
where $Z$ stands for partition function, and $\beta$ stands for
the thermodynamic term defined as $1/K_{B}T$. $K_{B}$ stands for
Boltzmann constant, and T stands for temperature.

In each Monte-Carlo sweep, we rand have been pickedomly pick one a
vector from lattice site $\vec{x}$, $\vec{A}[\vec{x}]$, and propose
a random move$.$ After updating this specific vector, we obtain a
updated configuration $\vec{A}'$, with associated probability $P(\vec{A'})=e^{-\beta H[\vec{A'}]}/Z$.
Now the relative probability of obtaining $A'$ vs $A$ is

\begin{equation}
r=\frac{P(\vec{A}')}{P(\vec{A})}=e^{-\beta(H[\vec{A}']-H[\vec{A}])}\equiv e^{-\beta(\Delta H)}
\end{equation}
where $\varDelta H=H[\vec{A}']-H[\vec{A}]$.

If $\varDelta H<0$, we always accept the change as our purpose is
to acquire the ground state of the system.

If $\varDelta H>0$, we accept the change with probability $r$.

The acceptance probability is calculated as following. We generate
a random number $x$ from $0$ to $1$ following uniform distribution.
If $x<r$, we accept the change. Otherwise, the change will not be
applied to the specified vector and the simulation carry on to the
next vector until we exhaust all the vectors in the lattice.

\begin{figure}[b]
\includegraphics[scale=0.3]{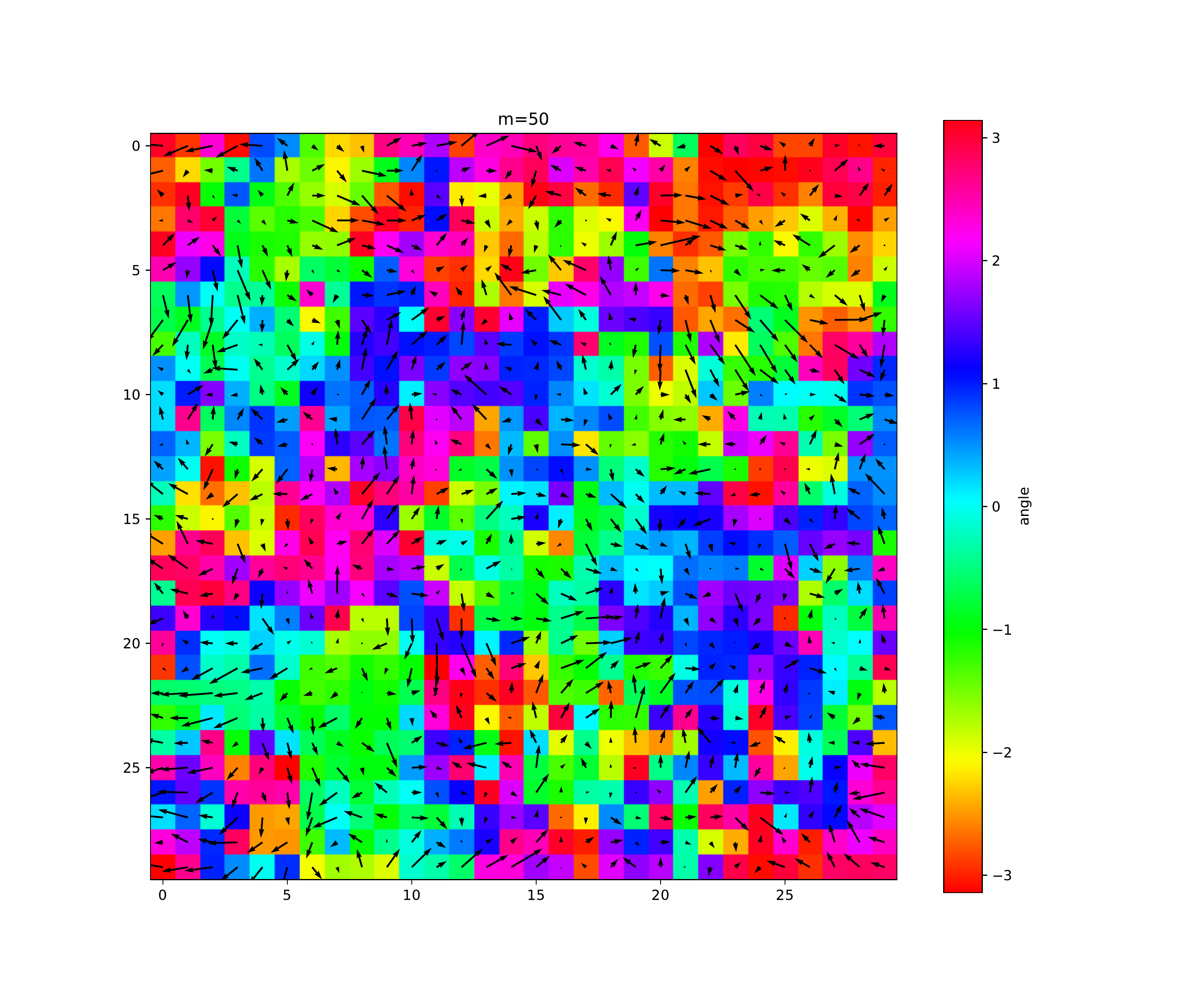}

\caption{\label{fig:mvf_config_example} Example configuration of massive vector
field simulated by Metropolis-Hasting algorithm. Color represents
angle of the vectors from $0$ to $2\pi$.}
\end{figure}

After a considerably large amount of Monte-Carlo sweep (see Appendix
\ref{sec:Simulation-parameters}), the vector field reaches equilibrium
and we obtain a relatively lowest energy configuration. The resultant
vector field is shown in Fig.\ref{fig:mvf_config_example}.

The arrows in the graph stands for our vectors in the vector field.
The length of every arrow represents the magnitude of the force of
that lattice site, and the direction of every arrow represents its
direction of force. The coloring in the graph visualize the direction
of every vector. Each color represents an angle value of the vectors
as the color scheme shows. Therefore, the graph can effectively visualize
the alignments between nearby vectors and serves as an efficient visual
aid for our simulations.

\section{Langevin equation in a vector field background\label{sec:Langevin+MVF}}

The massive vector field $\vec{A}$ we applied to our Langevin system
have random directions and lengths, which we use as model of the external
driving force $F_{ext}$ exerted on the Langevin particles $\vec{x}$. 

Take a particular realization of the massive vector field as the external
forces we can decompose 

\begin{equation}
F_{ext}(\vec{x})=\vec{A}(\vec{x})=\left(A_{h}(\vec{x}),\vec{A_{v}}(\vec{x})\right).
\end{equation}

Here, $\vec{A_{h}}(\vec{x})$ represents the horizontal force exerted
on the particles with respect to their positions. Similarly, $\vec{A_{v}}(\vec{x})$
represents the vertical force exerted on the particles with respect
to their positions. We can think of $A_{h}$ and $A_{v}$ as two independent
vector fields coupled through its kinetic and mass terms (Eqn.\ref{eq:mvf_hamiltonian}). 

Then we put the external forces (the effect of vector field) into
Langevin equation.
\begin{equation}
\gamma\dot{\vec{x}}=\text{\ensuremath{\vec{A}(\vec{x})}}+\eta(t)\label{eq:langevin3}
\end{equation}

We use uniform initial positions for the Langevin particles, and then
let them evolve under the Langevin equation Eq.\ref{eq:langevin3}. 

\begin{figure}
\includegraphics[scale=0.3]{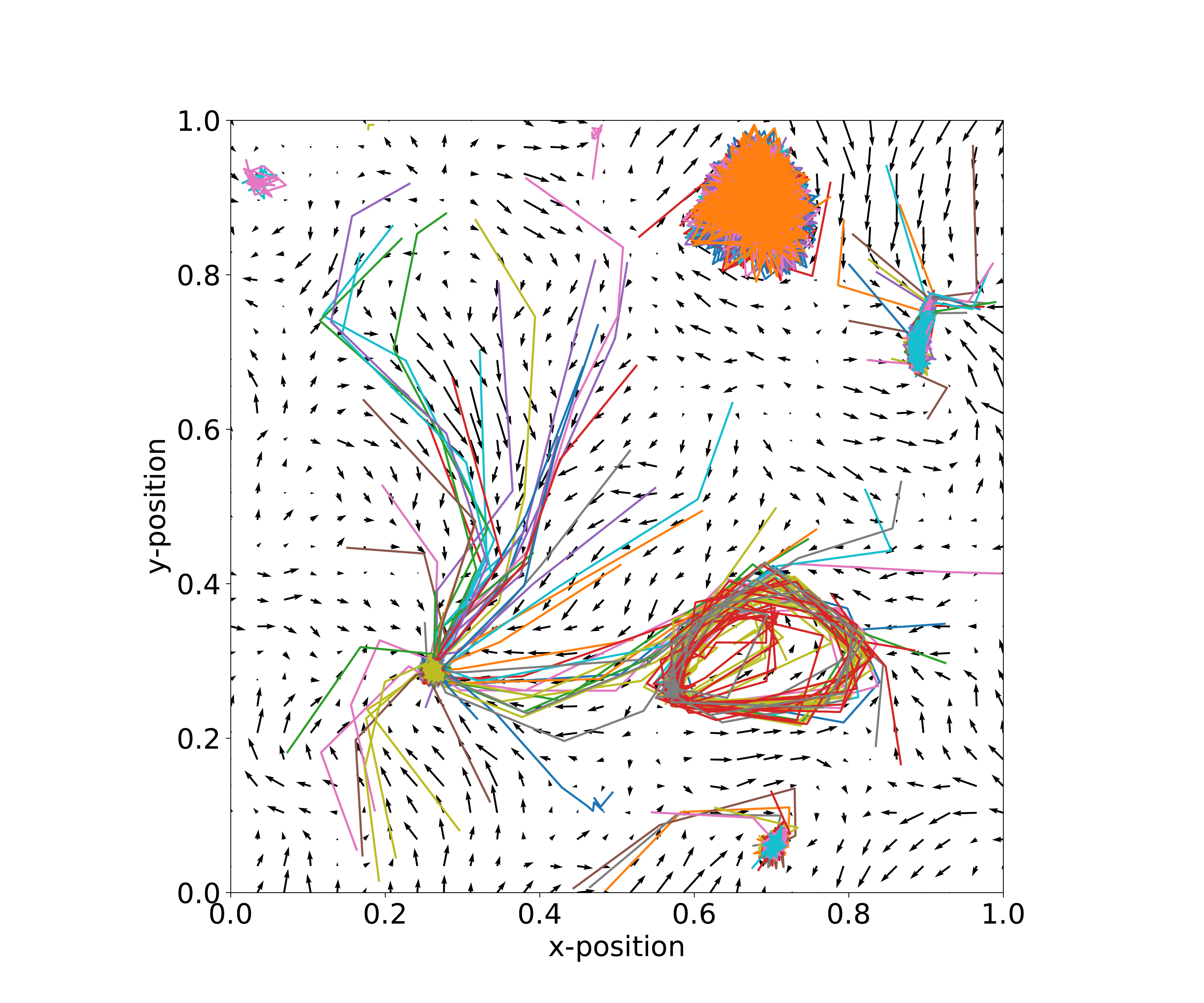}\caption{\label{fig:particle_trajectories} Example trajectories of particles
moving under Eq.\ref{eq:langevin3}. Different color represents different
initial conditions of the particles. }

\end{figure}

We plot the trajectories of particles under the driving of the vector
field, and color-coded them differently for different trajectories,
the result are shown in Fig.\ref{fig:particle_trajectories}.

When the particle passes through its trajectory, the kinetic energy
will be transferred into the system when the particle overcomes viscous
resistance, this energy is known as dissipation, the dissipation rate
is 

\begin{equation}
\varGamma(t)=\vec{F}_{diss}(t)\cdot\dot{\vec{x}}=\vec{A}(\vec{x})\cdot\dot{\vec{x}}(t)
\end{equation}

We can calculate the total dissipation energy from the particle to
the system by 

\begin{equation}
Diss=\intop\varGamma(t)dt=\vec{A}(\vec{x})\cdot\dot{\vec{x}}(t)dt\label{eq:Diss}
\end{equation}

Thus, the trajectory and dissipation of a particle are only affected
by the massive vector field and the random thermal motion. And then
we can plot the dissipation of each particle trajectory calculated
by Eq.\ref{eq:Diss} in the histogram.

Different combination of $m$ and $J$ results in different statistics
of the vector field, and in turn affects the typical trajectories
of the Langevin particles. Depending on the coordination of nearby
vectors, some trajectories, Langevin particles can either get trapped
into nearby attractors and limit cycles, or travel further from attractors
to attractors. 

\begin{figure}

\includegraphics[scale=0.3]{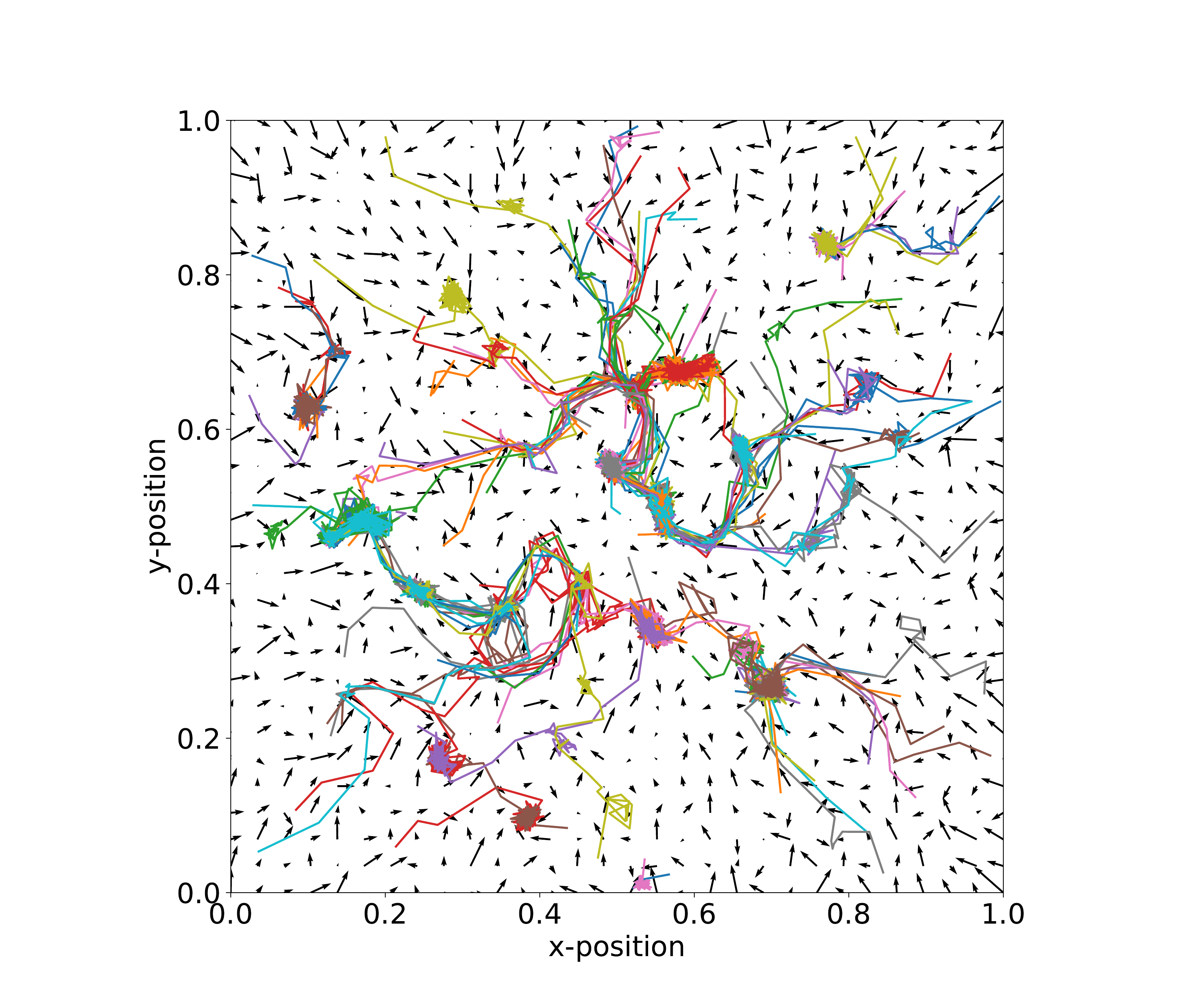}\caption{\label{fig:low_diss_traj} Example of a low dissipation trajectories
$(m=100,J=1)$}

\end{figure}

One situation is that $m$$\gg J$. In this situation, as shwon in
Fig.\ref{fig:low_diss_traj}, the distribution of the vector angles
is totally random. there is not any cooperation between nearby vectors.
All $A_{h}(\vec{x})$ vectors and $A_{v}(\vec{x})$ vectors are random
and approximately Gaussianly-distributed. The majority of the trajectories
get stuck at local attractors and seldom hop between nearby attractors.
We define this situation as low-dissipation condition.

\begin{figure}
\includegraphics[scale=0.3]{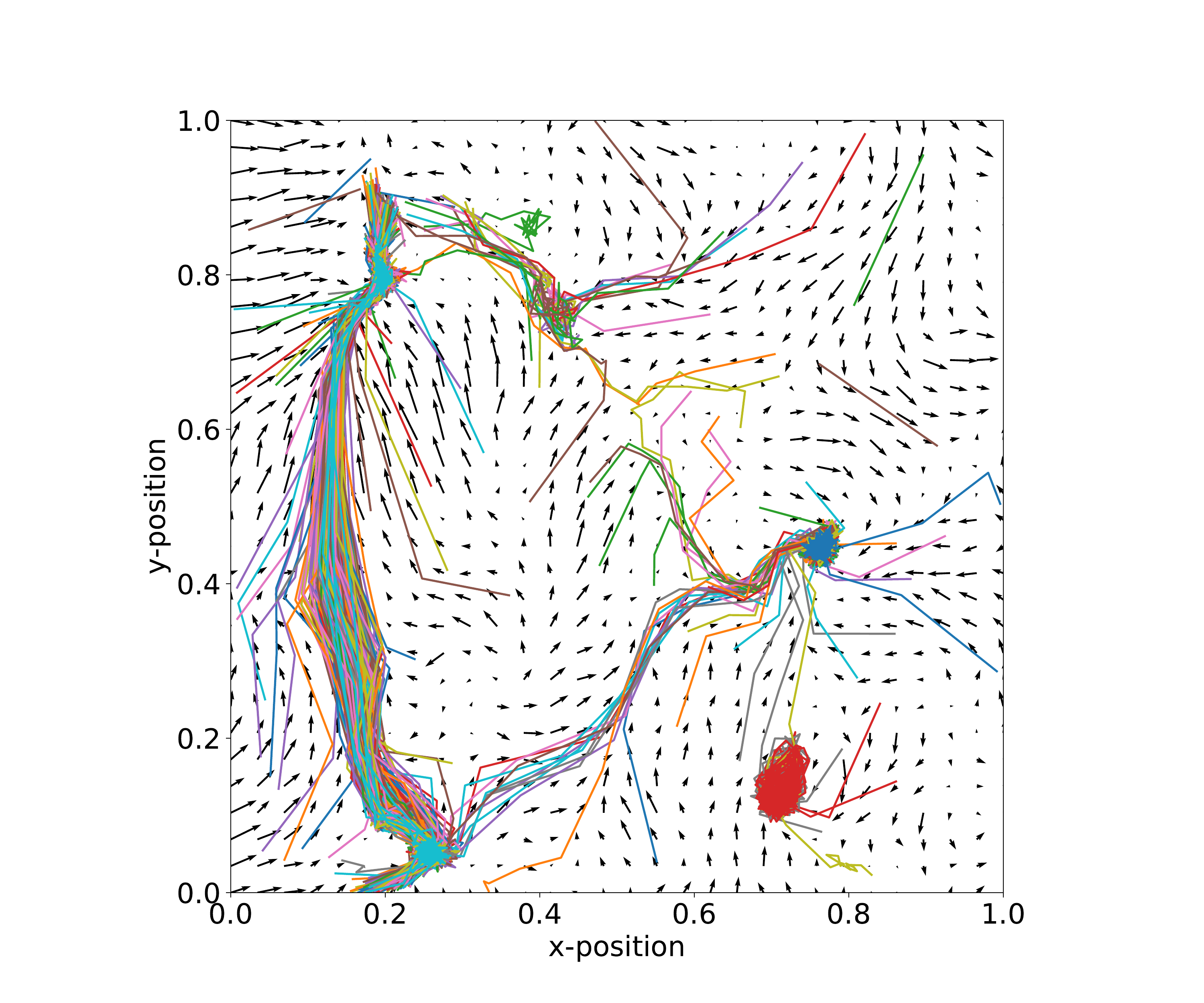}\caption{\label{fig:high_diss_traj} Example of trajectory in high dissipation
condition $(m=10,J=100)$}
\end{figure}

Another extreme situation is when $J\gg m$, and this corresponds
to total coordination, where the vector field resembles a constant
vector field. In this case, the external drive represents a constant
driving force. The resulting dynamics in this regime is not interesting.

Our ideal condition is that the when $m$ and $J$ are comparable.
In this situation, as shwon in Fig.\ref{fig:high_diss_hist} there
are some fix points that particles get trapped into. However, at the
same time, there are also extensive trajectories where particles can
hop between fix points.We define this situation as high-dissipation
condition.

In general, particles that have more extensive trajectories tend to
have higher dissipation, as evident in Eqn.\ref{eq:Diss} that total
dissipation energy is proportional to the total distance traveled
by the Langvein particle. Therefore, by observing the corresponding
dissipation energy we can infer the trajectories of the particles
in this system. Therefore, in the following we focus on the distribution
of dissipation energy. 

\section{Features of Low Dissipation Condition and High Dissipation Condition\label{sec:dissipation}}

\subsection{Compare Relatively Large Dissipation in Each Condition}

\begin{figure}
\includegraphics[scale=0.3]{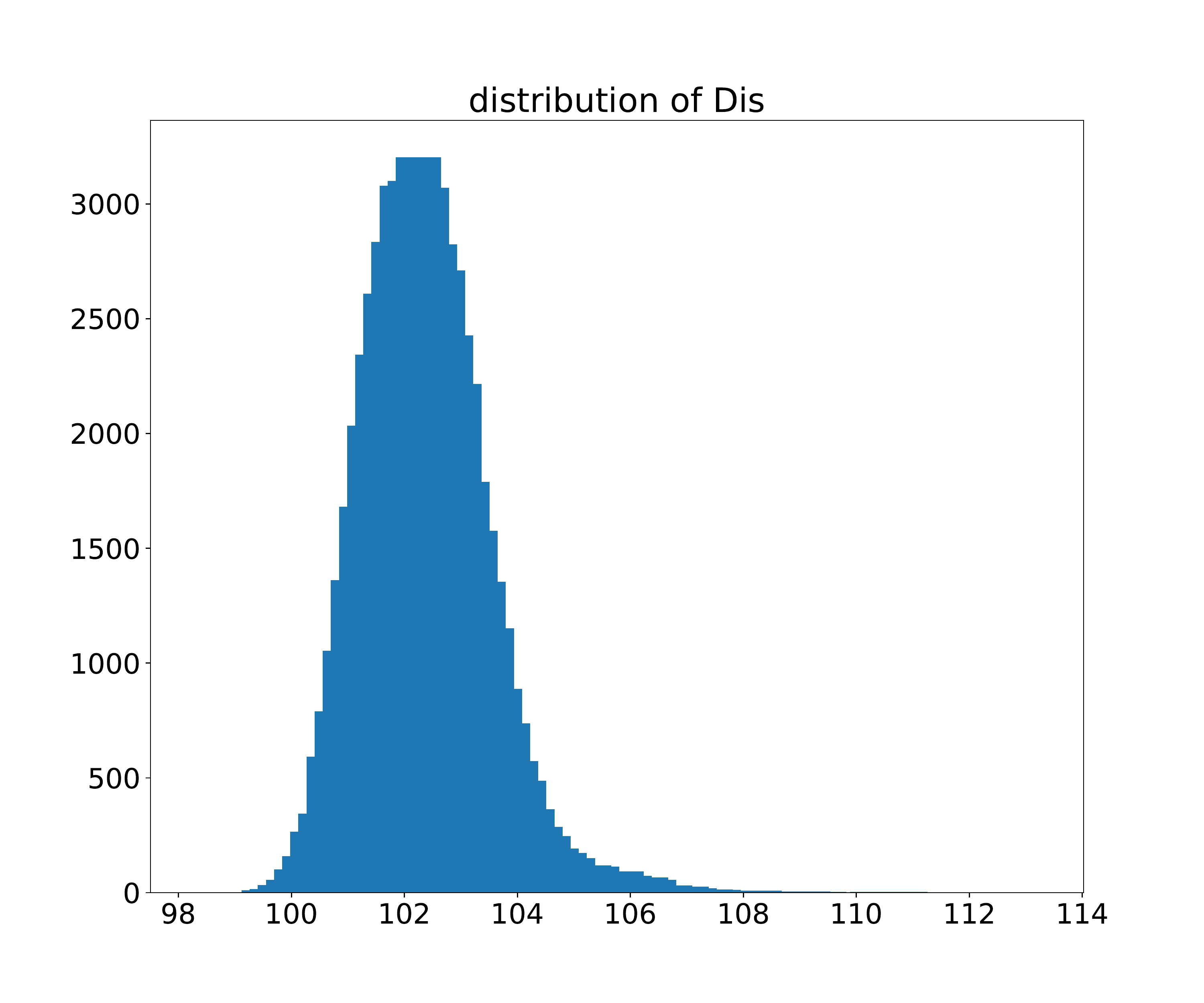}

\caption{\label{fig:low_diss_hist}Example of histogram of general dissipation
in low dissipation condition $m=100,J=1)$}
\end{figure}

In low-dissipation condition, the kinetic energy transfered into system
is generally small since the trajectories tend to get stuck. The distribution
of total dissipation of trajectories is shown in Fig.\ref{fig:low_diss_hist}.

\begin{figure}
\includegraphics[scale=0.3]{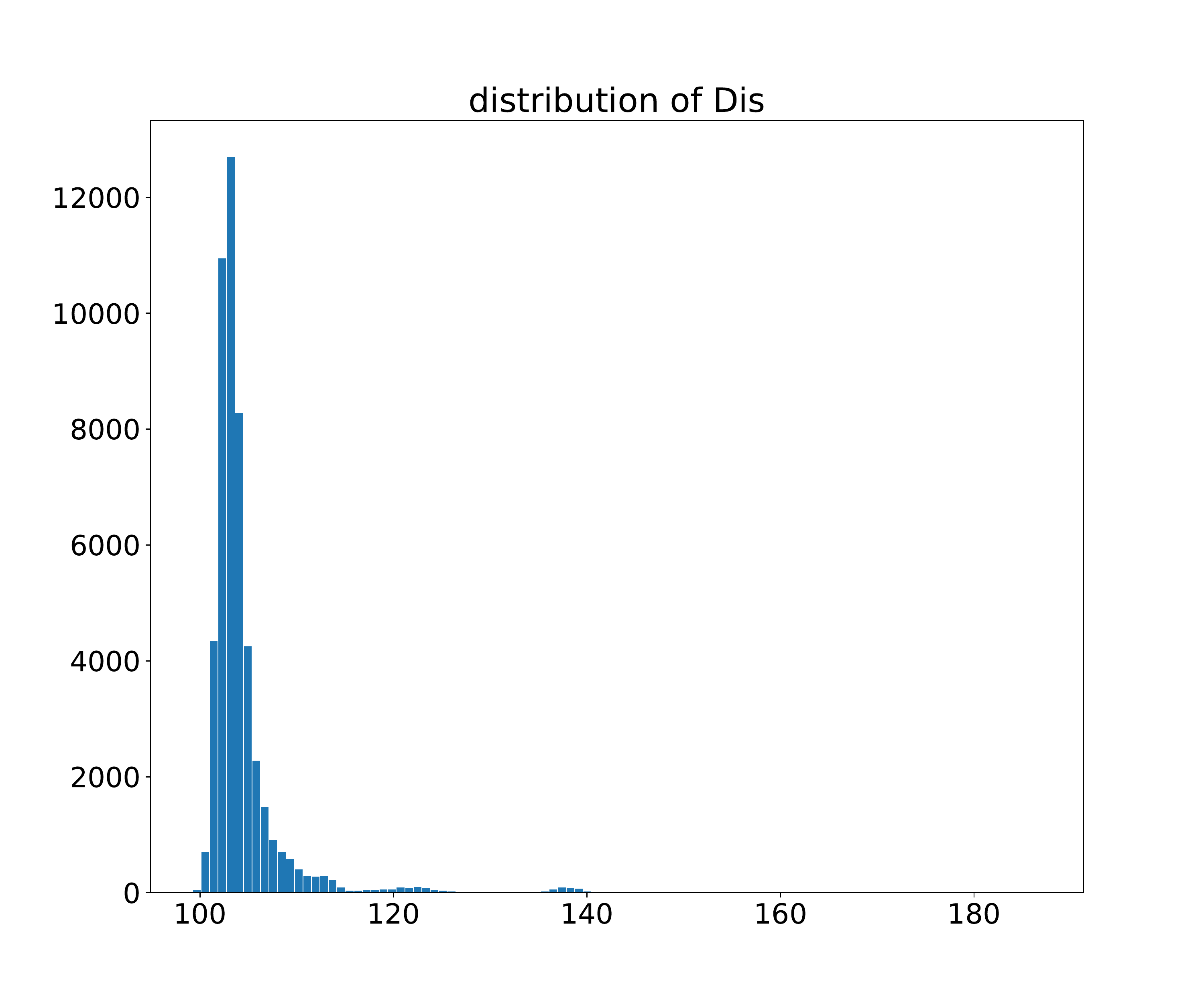}\caption{\label{fig:high_diss_hist}Example of histogram of general dissipation
in high dissipation condition $m=10,J=100)$}
\end{figure}

In high-dissipation condition, the total dissipation is larger than
in the low-dissipation condition, since particles can hop between
fix points before they finally stop. The distribution of total dissipation
is shown in Fig.\ref{fig:high_diss_hist}.

As what is shown in both Fig.\ref{fig:low_diss_hist} and Fig.\ref{fig:high_diss_hist}
that although there are some higher dissipation in each histogram,
their difference can't be detected obviously by comparing their mean
value. Therefore, we need another method to present the difference
between high dissipation phase and low dissipation phase. Our method
is to focus on the relatively high dissipation group, which corresponds
to the outlier part in the dissipation distribution. 

We start with calculating the Interquatile Range (IQR) of each distribution
by

\begin{equation}
IQR=Q3-Q1
\end{equation}

Here, IQR refers to the range between the $25^{th}$ percentile(Q1)
and the $75^{th}$ percentile(Q2) of each distribution. According
to the statistical rule of defining outliers by IQR, any data that
is larger than $Q3+1.5IQR$ is considered as outliers. Then, we can
calculate the mean value of dissipation from the outliers

\begin{equation}
\left\langle D_{outlier}\right\rangle =\underset{i}{\sum}D_{i}f_{i}\left[i\in\left(Q3+1.5IQR,max\right)\right],
\end{equation}

where $D_{i}$ represents the $i^{th}$ bin in the outlier dissipation
histogram and $f_{i}$ represents the frequency at this dissipation. 

In the same way, we can calculate the total dissipation by 
\begin{equation}
\left\langle D\right\rangle =\underset{j}{\sum}D_{j}f_{j}\left[j\in\left(min,max\right)\right]
\end{equation}

\begin{figure}
\includegraphics[scale=0.3]{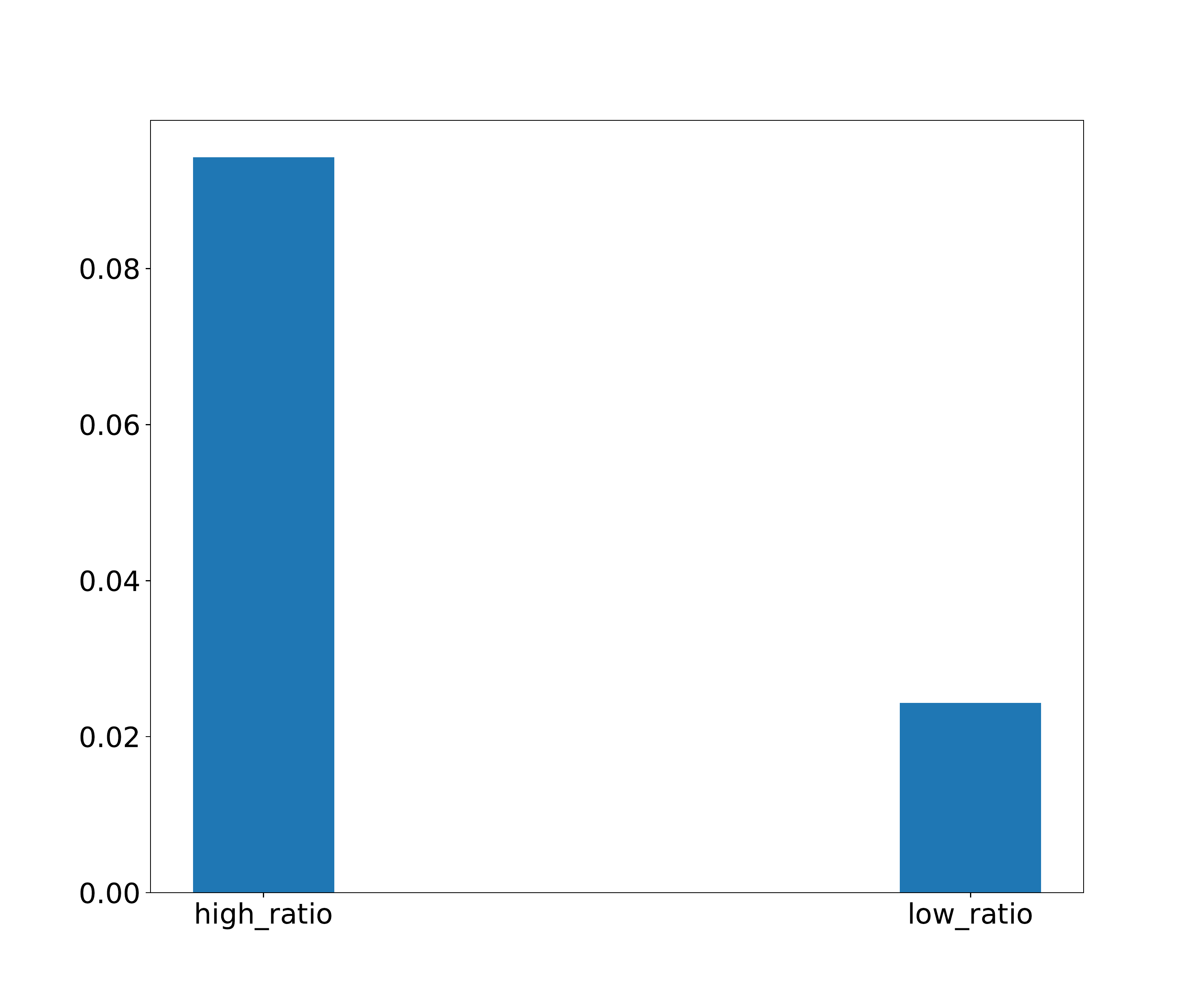}\caption{\label{fig:ratios} The two ratios of high dissipation phase and low
dissipation phase}
\end{figure}

With $\left\langle D_{outlier}\right\rangle $ and $\left\langle D\right\rangle $,
we can calculate their ratio by

\begin{equation}
Ratio=\frac{\left\langle D_{outlier}\right\rangle }{\left\langle D\right\rangle }
\end{equation}
The ratio stands for the portion of relatively large dissipation trajectories
from the overall dissipation. Hence we will have a more direct view
of the difference between the comparably high dissipation part of
trajectories in the two regimes. We plot this ratio in the high-dissipation
condition and low-dissipation condition in Fig.\ref{fig:ratios}. 

We can clearly see from the graph that the ratio of high dissipation
phase is greater than low dissipation phase. This means that the proportion
of large dissipation values is greater in high dissipation phase than
in the low dissipation phase, which provides an evidence of which
phase (low-dissipation vs high-dissipation) the system lives in. 

Also, we can use another method to provide evidence of high dissipation
condition. We can calculate the difference between the maximum value
of the dissipation and the value of $Q3+1.5IQR$ of the distribution,
which show the range of outlier in each distribution:

\begin{equation}
\Delta=D_{max}-(Q3+1.5IQR)
\end{equation}

\begin{figure}
\includegraphics[scale=0.3]{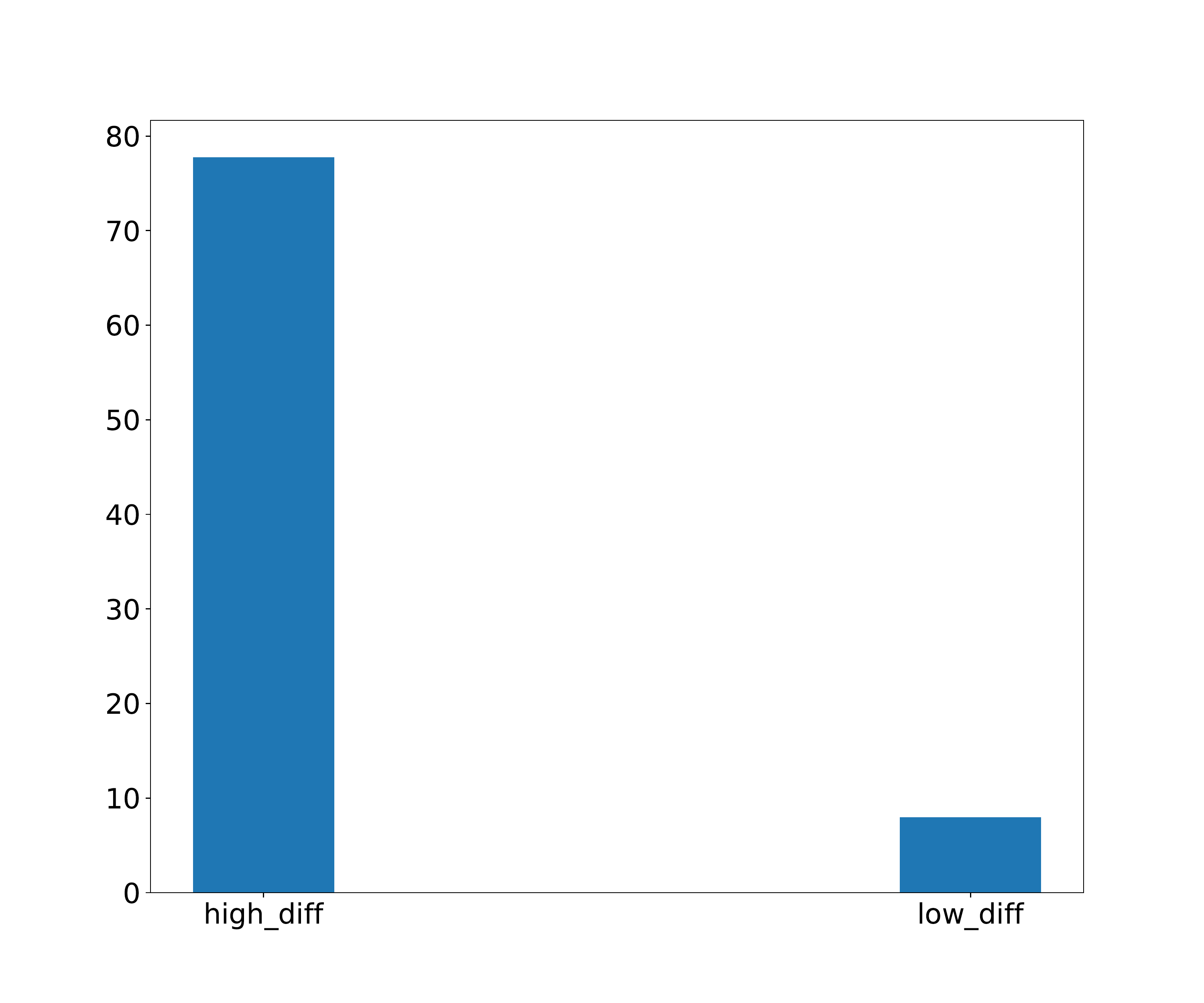}\caption{\label{fig:diff} The range of outliers in each dissipation condition. }
\end{figure}

We plot $\Delta$ in the two regimes in Fig.\ref{fig:diff}.

With $\Delta$ value, we believe that the greater range of the difference,
the larger dissipation values of the distribution has. Thus, the high-dissipation
phase has a greater value of $\Delta$ than the low-dissipation phase.

Overall, the high-dissipation and low-dissipation conditions in our system are markedly distinct. In both scenarios, Langevin particles are driven by an external force from the massive vector field and dissipate their kinetic energy into the environment. However, under the low-dissipation condition, since $m$, which measures the cooperativity of the vectors, is significantly larger than $J$, indicative of the vectors' randomness, the vector configuration tends to differ from that of neighboring vectors. Consequently, most particles quickly become trapped at various fixed points before traveling far, resulting in shorter trajectories and less kinetic energy being dissipated into the system. On the other hand, in the high-dissipation condition, $m$ and $J$ are of comparable magnitude. This means vectors exhibit some degree of cooperativity with their neighbors, allowing more particles to travel greater distances and hop between fixed points, thus creating extensive trajectories. In this condition, there is a higher level of dissipation, particularly from particles that travel further than most others. This distinction is illustrated by the proportions (ratios) and ranges of relatively high dissipation for each condition, as shown in Fig.\ref{fig:ratios} and Fig.\ref{fig:diff}. In both figures, the range or ratio for the high dissipation condition is greater than that of the low dissipation condition. In conclusion, we can use both $Ratio$ and $\Delta$ as reliable indicators to determine whether the system is in the localized or delocalized phase.

\section{Conclusion}

In this work, we integrate the Langevin model with a massive vector field to simulate the behavior of random walks out of equilibrium under random drives. Our primary focus is on the particle trajectories and their energy dissipation, which occurs as the particles' kinetic energy is transferred to the environment while overcoming drag forces. Our numerical analysis confirms the existence of two distinct phases in our model: the low dissipation phase and the high dissipation phase. We introduce two metrics to diagnose these phases. We believe that our identification and characterization of these two dynamical phases will provide fundamental insights for understanding and manipulating nonequilibrium phases of matter. Looking ahead, we aim to investigate the scaling properties of dissipation within this model. We are also interested in exploring the relationship between dissipation and other thermodynamic quantities, such as configuration entropy. Ultimately, uncovering the complete phase diagram of the model represents a crucial future direction for our research.

\begin{acknowledgments}
Z.L. and J.Y. acknowledges support from Zhixin High School Science
Outreach Program. Z.L. and J.Y. would like to thank Weishun Zhong
for suggesting this project, and guidance throughout carrying out
the simulation and preparing this manuscipt. 
\end{acknowledgments}

\appendix

\section{Simulation parameters\label{sec:Simulation-parameters}}

In this section, we detail the parameter values used in the simulation.
In our simulation, we discretize the lattice site of the massive vector
field in Eqn.\ref{eq:mvf_hamiltonian} into 30 sites along each axies.
The temperature of vector field in the Metropolis algorithm is set
to be $0.15$. The temperature of the Langevin system is $0.001$.
The simulation time for massive vector field is 500 Monte-Carlo sweeps.
We choose the time interval of discretizing the Langevin eqaution
as 0.02 time units, and the total duration of the simulation is 500
time units.The drag coefficient in Eqn.\ref{eq:langevin3} is set
to be $\gamma$=$20$. The size of the periodic box in the Langevin
system is set to be $1$. In all the simulations, we average over
1000 different initial conditions of the Langevin particles in each
realization of the vector field, and average over 50 different realizations
of the vector field to obtain the statistics presented in the main
text. 

.

\bibliographystyle{unsrt}
\bibliography{refs.bib}

\end{document}